\documentclass{aastex}
\usepackage{emulateapj5,epsfig,onecolfloat,apjfonts}   

\shorttitle{Unresolved X-ray Sources in Clusters of Galaxies}
\shortauthors{S.~M.~Molnar, J.~P.~Hughes, M.~Donahue and M.~Joy}

\def\CHANDRA{{\it Chandra }}
\def\ROSAT{{\it ROSAT }}
\def\flux{\rm\;erg\;s^{-1}\;cm^{-2}}
\def\lum{\rm\;erg\;s^{-1}}
\def\logn{$\log N - \log S\;$}

\newenvironment{onecolumnfigure}{
\def\@captype{figure}
\noindent\begin{minipage}{0.9999\linewidth}\begin{center}}
{\end{center}\end{minipage}\smallskip}

\begin{document}

\twocolumn
[
\title{{\it CHANDRA} Observations of Unresolved X-ray Sources around two Clusters of Galaxies}

\author{
Sandor M. Molnar\altaffilmark{1}, John P. Hughes\altaffilmark{1}, 
Megan Donahue\altaffilmark{2} and Marshall Joy\altaffilmark{3}
}

\begin{abstract}
We have searched for unresolved X-ray sources in the vicinity of two 
rich clusters of galaxies: Abell 1995 (A1995) and MS 0451.6-0305 (MS0451), 
using the \CHANDRA X-ray observatory.
We detected significantly more unresolved sources around A1995 
than expected based on the number of X-ray sources to the 
same flux limit detected in deep \CHANDRA observations of blank fields.
Previous studies have also found excess X-ray sources in the 
vicinity of several nearby clusters of galaxies using \ROSAT$ $, and 
recently in more distant ($z \approx 0.5$) clusters (RXJ0030 and 3C295)
using \CHANDRA. 
In contrast, we detect only 14 unresolved X-ray sources near MS0451, 
which is consistent with the number expected from a cluster-free background. 
We determine the luminosity functions of the extra sources under the 
assumption that they are at the distance of their respective clusters.
The characteristic luminosity of the extra sources around 
A1995 must be an order of magnitude fainter than that of the extra 
sources around RXJ0030 and 3C295. 
The apparent lack of extra sources around MS0451 is consistent with its
greater distance and the same characteristic luminosity as the A1995 sources. 
Hardness ratios suggest that, on average, 
the extra sources in A1995 may have harder spectra than those of RXJ0030 and 3C295. 
These results indicate that different classes of objects may dominate in 
different clusters, perhaps depending on the formation history and/or dynamical 
state of the accompanying cluster.
\end{abstract}

]

\altaffiltext{1}{Department of Physics and Astronomy, Rutgers University, 
136 Frelinghuysen Road, Piscataway, NJ~08854}

\altaffiltext{2}{Space Telescope Science Institute, 3700 San Martin Drive, 
Baltimore, MD 21218}

\altaffiltext{3}{Department of Space Science, NASA, Marshall Space Flight Center,
Huntsville, Al 35812}

\section{Introduction}
\label{s:Introduction}

Evidence has accumulated recently that there are more X-ray point sources
in the direction of clusters of galaxies than toward cluster-free regions of the sky. 
Henry \& Briel (1991), using {\it ROSAT} PSPC observations, found just about twice 
as many unresolved sources around Abell 2256 (at $z = 0.06$, Struble \& Rood 1991) 
as expected from blank field (no clusters) background observations. 
The luminosity of these sources, assuming they are at the redshift of 
the cluster was found to be about $10^{42} \lum$ or greater (in 0.5 - 2 keV).
These sources have high X-ray to optical flux ratios.
Some of the sources in A2256 were identified as cluster member galaxies.
Henry and Briel also discuss the possibility that the emission from these
sources is due to hot gas in galaxies not removed by ram pressure 
or evaporation, or due to shocks in gas from merging.

Similarly, Lazzati et al. (1998), analyzing ROSAT PSPC images, found an excess 
number of unresolved X-ray sources in the fields of two nearby clusters: A194 and 
A1367 ($z =$ 0.018 and 0.022). 
The spectra of the sources were consistent with thermal bremsstrahlung with 
$T \le 2$ keV.
Lazzati et al. also found evidence for association of some of these sources 
with cluster member galaxies, implying luminosities between 
0.6 and 6.6 $\times 10^{41}\lum$ in the 0.5 - 2 keV band.
X-ray emission from hot gas associated with cluster member galaxies had been 
reported earlier, also based on ROSAT PSPC observations 
(Grebenev et al. 1995; Bechtold et al. 1983).

Indirect evidence has also been presented for the existence of 
an excess population of unresolved X-ray sources associated with
Abell clusters.
Soltan \& Fabricant (1990) using Imaging Proportional Counter data 
from the {\it Einstein Observatory} found excess fluctuations in nearby 
galaxy clusters which could be explained by assuming the presence of 
low luminosity sources ($\approx 4 \times 10^{41} \lum$) in clusters with 
extent less than 1$\arcmin$.
They discuss the possibility that the emission from these sources is due to
low luminosity AGNs, or to hot gas in member galaxies. 
Soltan et al. (1996) found a correlation between the surface brightness of the
X-ray background and Abell clusters on scales of a degree, 
which is much larger than the X-ray emission from the intracluster gas.
The characteristic length was found to be about 10 $h^{-1}$ Mpc in radius 
(where $H_0 = h 100 \;\rm Km\;s^{-1}\;Mpc^{-1}$), 
i.e. extra X-ray emission was found around Abell clusters out to about 
15 Mpc ($h$ = 0.65).
Soltan et al. could not explain the extra X-ray emission based on known 
sources, or random fluctuations in their number density. 
They estimated the required number of excess sources to be about 50$\%$ 
above the expected number of background sources.

Most recently Cappi et al. (2001), using \CHANDRA ACIS 
(Advanced CCD Imaging Spectrometer) observations, found twice as many unresolved 
X-ray sources in the images of two distant clusters of galaxies, 
3C295 ($z = 0.46$ Dressler \& Gunn 1992), and RX J003033.2+261819 
(RXJ0030; $z = 0.5$ Vikhlinin et al. 1998), 
as expected from a cluster-free background to their flux limits
(Giacconi et al. 2001; Mushotzky et al. 2000).

Our main goal in this letter is to present new results on the flux and number 
distributions of unresolved X-ray sources based on our \CHANDRA ACIS
observations of two rich clusters of galaxies: A1995 and MS0451. 
We also address briefly the nature of the excess sources.

\begin{table}[t]
\footnotesize
\begin{center}
\caption{Unresolved X-ray sources detected in A1995.}
\begin{tabular}{lccc}
               &            &           &        \\
\hline \hline
 X-RAY SOURCE  &   COUNTS   &  COUNTS   & R$^a$  \\
               & 0.5-2 KEV  & 2-10 KEV  & Mag    \\
\hline
CXOU J45308.70+580313.8 & 585$\pm$25 & 179 $\pm$14 & 18.8 \\
CXOU J45305.82+580309.1 & 342$\pm$19 & 79.1 $\pm$9.6 & 19.2 \\
CXOU J45307.10+580205.8 & 306$\pm$18 & 74.2 $\pm$9.2 & 20.8 \\
CXOU J45327.65+580339.5 & 166$\pm$14 & 39.7 $\pm$7.6 & 20.9 \\
CXOU J45305.45+580033.9$^b$ & 69.5$\pm$8.5 & $<$10         & 10.2 \\
CXOU J45246.31+580059.7 & 65.4$\pm$8.2 & 14.5 $\pm$4.0 & 20.9 \\
CXOU J45317.46+580003.0 & 56.3$\pm$8.2 & 20.9 $\pm$5.9 & 20.3 \\
CXOU J45233.22+580559.2 & 48.8$\pm$7.1 & 17.6 $\pm$4.4 & 19.5 \\
CXOU J45229.56+580418.2 & 37.5$\pm$6.2 & 20.4 $\pm$4.7 & $>$22 \\
CXOU J45230.74+580448.5$^c$ & 34.5$\pm$6.0 & $<$10           & 16.0 \\
CXOU J45233.56+580456.3 & 28.4$\pm$5.4 & 23.0 $\pm$5.0 & 22.0 \\
CXOU J45324.68+580318.4 & 26.7$\pm$5.7 & $<$10           & $>$22 \\
CXOU J45316.75+575928.6 & 26.6$\pm$6.1 & $<$10           & $>$22 \\
CXOU J45248.65+580255.5$^b$ & 24.3$\pm$5.3 & $<$10           & 13.2 \\
CXOU J45301.83+580005.7 & 22.1$\pm$5.0 & 11.4 $\pm$4.0 & $>$22 \\
CXOU J45255.19+580056.0 & 20.7$\pm$4.7 & $<$10           & $>$22 \\
CXOU J45244.00+580203.3 & 20.6$\pm$4.7 & 14.2 $\pm$4.0 & $>$22 \\
CXOU J45315.41+580448.5 & 20.5$\pm$5.6 & $<$10           & 21.0 \\
CXOU J45253.64+580020.3 & 20.1$\pm$4.6 & $<$10           & $>$22 \\
CXOU J45319.11+580134.4 & 17.2$\pm$4.8 & $<$10           & $>$22 \\
CXOU J45313.22+580126.9 & 16.5$\pm$4.6 & $<$10           & $>$22 \\
CXOU J45315.55+580117.4 & 16.3$\pm$4.9 & $<$10           & 20.3 \\
CXOU J45242.50+580159.1 & 15.8$\pm$4.1 & $<$10           & $>$22 \\
CXOU J45228.20+575954.5 & 15.0$\pm$4.0 & 7.2 $\pm$3.0 & 21.1 \\
CXOU J45231.03+580010.9 & 14.5$\pm$3.9 & 5.1 $\pm$2.4 & 21.1 \\
CXOU J45235.73+580656.1 & 12.8$\pm$3.7 & $<$10           & $>$22 \\
CXOU J45245.50+580519.5 & 12.2$\pm$3.6 & $<$10           & $>$22 \\
CXOU J45234.49+575904.0 & 11.9$\pm$3.6 & 14.2 $\pm$4.2 & $>$22 \\
CXOU J45251.93+580046.2 & 10.1$\pm$3.3 & 12.3 $\pm$3.7 & 20.9 \\
CXOU J45322.04+575858.7 & $<$10           & 25.3 $\pm$6.4 & $>$22 \\  
\hline\hline
\end{tabular}
\end{center}
$^a$ R band magnitude of optical counterparts/limit if not detected \\
$^b$ GSC2 object (\texttt{http://www-gsss.stsci.edu/gsc/gsc2})      \\
$^c$ IRAS Galaxy (F14511+5816)                                      \\
\end{table}

\section{Data Processing and Analysis}
\label{s:Data}

A1995 is a rich cluster at $z$ = 0.32 with an intracluster gas temperature 
of $T_X = 7.6$ keV (Patel et al. 2000). 
A1995 was observed with the \CHANDRA ACIS 
in May and July 2000 for 35 k sec and 12 k sec. The aim point was on 
the back-illuminated chip S3 of ACIS-S.
The data were taken in full frame mode with a readout time of
3.2 sec. We used standard \CHANDRA software tools to
clean the data of time intervals with high background and/or
bad aspect, remove bad or flickering pixels, and correct for
event gains for a focal plane temperature of $-$120 C.
The final effective exposure time of the merged data after cleaning
was 54.5 k sec (for details see Joy et al. 2002).
MS0451 is a rich cluster at $z$ = 0.55, with an intracluster gas temperature 
of $T_X = 10.9 \pm1.2$ keV (Donahue et al. 1999).
MS0451 was observed with \CHANDRA in October 2000 for 45 k sec, also on the 
back-illuminated chip S3. After similar data processing, the cleaned data set 
has a 44.75 k sec exposure time (for details see Donahue et al. 2002).

We analyzed data from within $\le 5 \arcmin$ of the nominal center of 
the clusters, close to the optical axis, where the point spread 
function (PSF) is not degraded significantly, and the change in effective 
area is negligible.
We made images in the $0.5-2$ keV and $2-10$ keV energy bands, 
where the calibration is most reliable.
We used the \texttt{wavedetect} package of the \CHANDRA Interactive Analysis
of Observations setting the probability of fake detection to 1$\times 10^{-6}$ 
(corresponding to 4.7 $\sigma$, for details, see Cappi et al. 2001).
We kept sources which had a signal-to noise ratio of 3 (using a local background estimate)
corresponding to the detection of about 10 counts within the source area.
On average, we would expect one spurious source in an area of the sky 
four times that of the S3 chip.
These are similar to the conditions used by Cappi et al. (2001).
The background cluster emission is very smooth, and, according to 
our Monte Carlo simulations, the probability of the cluster emission 
resulting in false detection of a point source is negligible. 
We used a power law with a photon index of $\Gamma = 1.4$
when converting count rates to fluxes.
We did not correct for vignetting since in this soft band (0.5 - 2 keV)
and off axis angles of $\le 5\arcmin$ the correction is negligible.
Note, that in this soft band the fluxes have only a weak dependence on
the power law slope and absorbing column density, which are 
$N_{\rm H}$ = 3.5 $\times 10^{20}\;\rm cm^{-2}$ and 
3.9 $\times 10^{20}\;\rm cm^{-2}$ for A1995 and MS0451.

\begin{table}[t]
\footnotesize
\begin{center}
\caption{Unresolved X-ray sources detected in MS0451.}
\begin{tabular}{lcc}
               &             &          \\
\hline \hline
  X-RAY SOURCE &  COUNTS     & COUNTS   \\
               & 0.5-2 KEV   & 2-10 KEV \\
\hline
 CXOU J45419.63$-$30420.5$^{a,b}$ & 746 $\pm$28     & 150   $\pm$13   \\
 CXOU J45356.32$-$25837.7$^a$ & 396 $\pm$20     & 121   $\pm$11   \\
 CXOU J45422.59$-$30035.2 &  87.9 $\pm$ 9.5 &  17.8 $\pm$4.4  \\
 CXOU J45424.75$-$25849.8 &  69.0 $\pm$ 8.5 &  33.1 $\pm$6.0  \\
 CXOU J45426.07$-$30013.2 &  57.1 $\pm$ 7.7 &  12.1 $\pm$3.7  \\
 CXOU J45412.81$-$30047.7 &  45.2 $\pm$ 8.1 &     $<$10  \\
 CXOU J45419.20$-$30521.2 &  28.4 $\pm$ 6.0 &     $<$10  \\
 CXOU J45408.57$-$30521.2 &  27.4 $\pm$ 6.0 &     $<$10  \\
 CXOU J45410.88$-$30125.2 &  21.4 $\pm$ 5.7 &     $<$10  \\
 CXOU J45355.65$-$30409.6$^b$  &  19.7 $\pm$ 4.9 &     $<$10  \\
 CXOU J45406.70$-$30412.3$^{a,b}$  &  18.0 $\pm$ 4.6 &     $<$10  \\
 CXOU J45421.95$-$25816.2 &  17.2 $\pm$ 4.2 &  43.9 $\pm$6.8 \\
 CXOU J45404.19$-$30403.7 &  12.0 $\pm$ 3.9 &     $<$10  \\
 CXOU J45421.39$-$30132.4 &  10.9 $\pm$ 3.5 &     $<$10  \\
 CXOU J45356.73$-$30226.1 & $<$10 &  23.1 $\pm$5.6 \\  
\hline\hline
\end{tabular}
\end{center}
$^1$ GSC2 object  \\
$^2$ 2MASS source (\texttt{http://www.ipac.caltech.edu/2mass})\\
\end{table}

\section{Results}
\label{s:Results}

We detected 29 and 14 unresolved X-ray sources in the fields of A1995 and MS0451.
The sources show no correlation with the spatial distribution of the cluster emission.
Source details are given in Table 1 and 2 and 
the \logn curves are plotted in Figure~\ref{F:FIG1}.
In Figure 1 we also show the \logn curves for the sources near RXJ0030 and 3C295 
(Cappi et al. 2001).
We quote source densities in terms of number per ACIS chip ($8\arcmin \times 8\arcmin$) 
in order to show the actual numbers of unresolved sources as detected.
The expected \logn curves from cluster-free background (dashed lines)
is taken from Mushotzky et al. (2000) which predicts a slightly higher 
number density of background sources than Rosati et al. (2002), 
Campana et al. (2001) and Giacconi et al. (2001), and slightly less than 
Brandt et al. (2001). 
We choose the background predicted by Mushotzky et al. (2000) since it
provides the best fit to the high flux end of our \logn curve.
We detect sources with fluxes brighter than $6~\times~10^{-16} \flux$ and 
$8~\times~10^{-16} \flux$ in A1995 and MS0451.
We obtained $R$ band magnitudes for unresolved X-ray sources in A1995
(see details in Patel et al. 2000).
The 0.5-2 keV X-ray to $R$ band optical flux ratios 
of unresolved sources in the field of A1995 (cf. Table 1) seem to have 
a similar distribution to those of background sources (Mushotzky et al. 2000).

The \logn curve of the unresolved sources in the A1995 field is steeper 
than that of the background between 1 and $3 \times 10^{-15} \flux$, 
which indicates a build up of extra sources (excess number of sources
relative to the background) with fluxes in this interval. 
There is an indication that for fluxes less than $10^{-15} \flux$ the slope
of the \logn curve of A1995 is close to that of the background, suggesting
that there are no extra sources with flux below this value. 
The \logn curves of unresolved sources in 3C295 and RXJ0030 show similar 
signs of a cut off at low fluxes in the distribution of the excess sources 
(cf. Figure~\ref{F:FIG1}). 
Overall we find 29 unresolved sources in the A1995 field, as opposed to the 
expected number for a cluster free background of about 17.

In the field of our more distant cluster, MS0451, we find 14 unresolved sources, 
which is within 1$\sigma$ from the number expected based on a cluster free background.
Since MS0451 is at about the same redshift as RXJ0030 and 3C295,
our exposure time is longer than those of RXJ0030 and 3C295, 
and we found no extra sources in the MS0451 field, we have the simple and
potentially important finding that: 
{\it not all} clusters have extra unresolved sources associated with them
at the flux limits of these \CHANDRA observations.

\begin{onecolumnfigure}
{\epsfxsize=8.8cm\epsffile{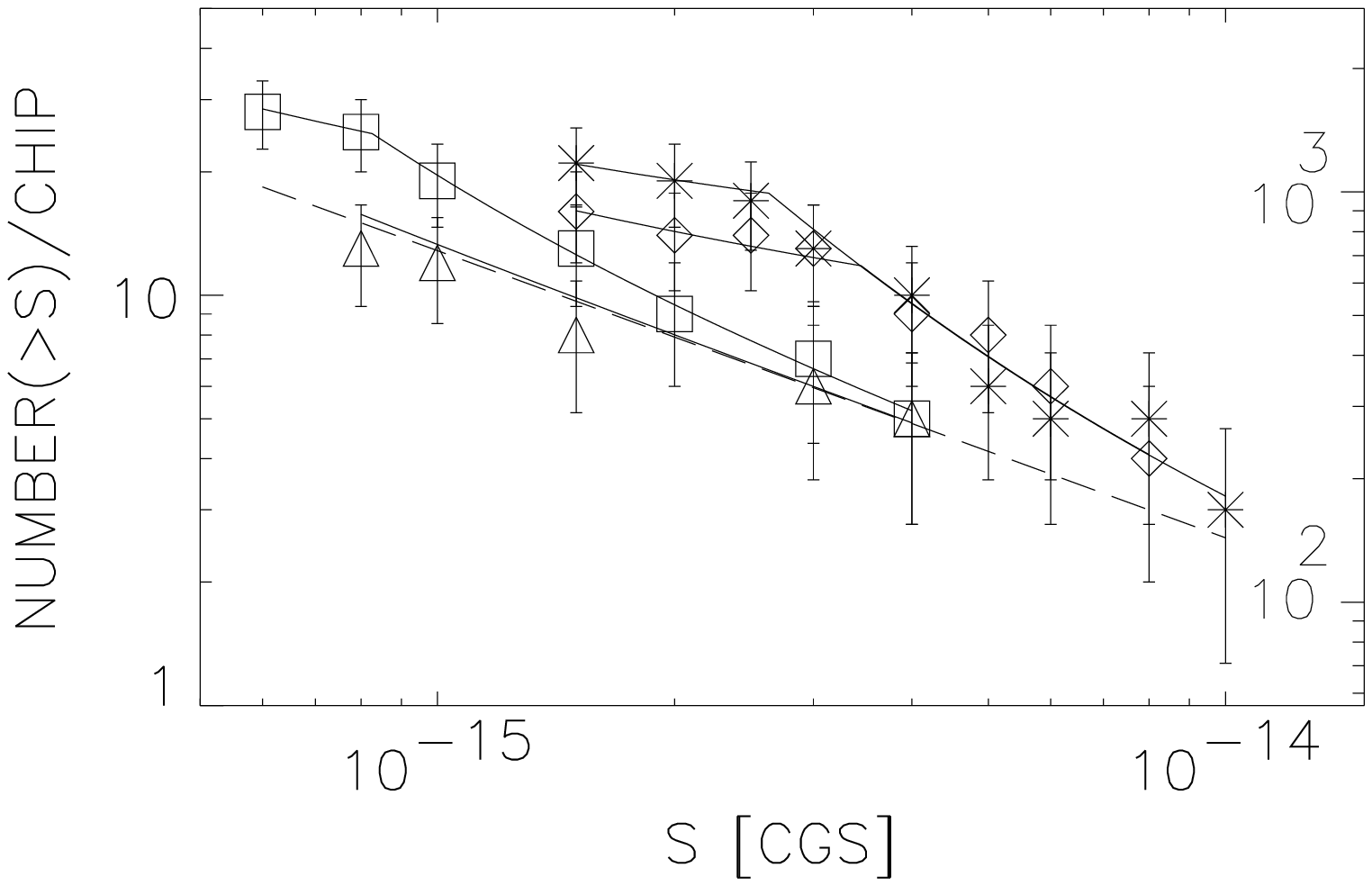}}
\figcaption{ \label{F:FIG1}  
$\log N(\ge S)$--$\log S$ per ACIS chip ($8\arcmin \times 8\arcmin$)
in the 0.5-2.0 keV band for unresolved X-ray sources in the fields of 
Abell 1995 and MS0451 (squares and triangles, our results), and
RXJ0030 and 3C295 (stars and diamonds, Cappi et al 2001). 
The expected background is shown with dashed line. 
The expected background plus cluster sources, using their derived luminosity 
functions (see text), are shown using solid lines.
On the left Y axis we show the corresponding numbers per square degree for comparison.
}
\end{onecolumnfigure}

In this letter we estimate the luminosity functions for the extra sources 
in A1995, RXJ0030 and 3C295 using their \logn curves, assuming the excess 
sources are at the redshifts of their respective clusters.
As customary, we used a luminosity function of the form 
$\Phi(L) \propto (L/L_*)^{-\alpha}$ with a lower cut off at a 
characteristic luminosity, $L_*$ (i.e. there are no sources with $L < L_*$). 
The fits, which are a good description of the \logn distribution, 
are shown using solid lines in Figure~\ref{F:FIG1}.
We found that the same slope, $\alpha = 3.1$, could describe the unresolved 
sources in all three clusters. 
However, the characteristic luminosities were found to be 
$L_*$= 0.5, 4.0, and 4.8 $\times 10^{42}\;\lum$ in A1995, RXJ0030 and 3C295. 
The total number of extra sources based on these fits are: 10, 12, and 7 
(with about $\pm$3 statistical error) for A1995, RXJ0030 and 3C295.
Due to the cluster emission, our ability to detect faint sources
decreases toward the center (about 20 counts per detect cell for
A1995). This effect is a fraction of the Poisson fluctuations
(it corresponds to missing at most a single additional source) and
so we ignore it.
We verified, by means of Monte Carlo simulations, that the observations can 
be drawn from the assumed background distribution plus a distribution based 
on the derived luminosity functions of the extra cluster sources, 
and that the observations can not be explained using the same luminosity 
function for all clusters.
Although the uncertainties in the luminosity function parameters are quite high: 
$\pm 0.5$ in the slope, and 30$\%$-40$\%$ in the lower cut-off, 
the characteristic luminosity of unresolved sources associated with A1995 are 
{\it significantly} (about one order of magnitude) less than those 
associated with RXJ0030 and 3C295.
Furthermore, when the luminosity function of unresolved sources near A1995
scaled to the greater distance of MS0451 the resulting \logn is fully 
consistent with the observed one (see Figure~\ref{F:FIG1}, solid curve 
associated with MS0451).
On the other hand, it is not consistent with the scaled luminosity function
of either RXJ0030 or 3C295 (which are roughly the same distance as MS0451).
We expect the effect of spatial variations in the PSF, and effective area, 
on the completeness function to be small, with an overall effect on the 
derived luminosity function much less than the errors quoted above.

\begin{onecolumnfigure}
{\epsfxsize=8.5cm\epsffile{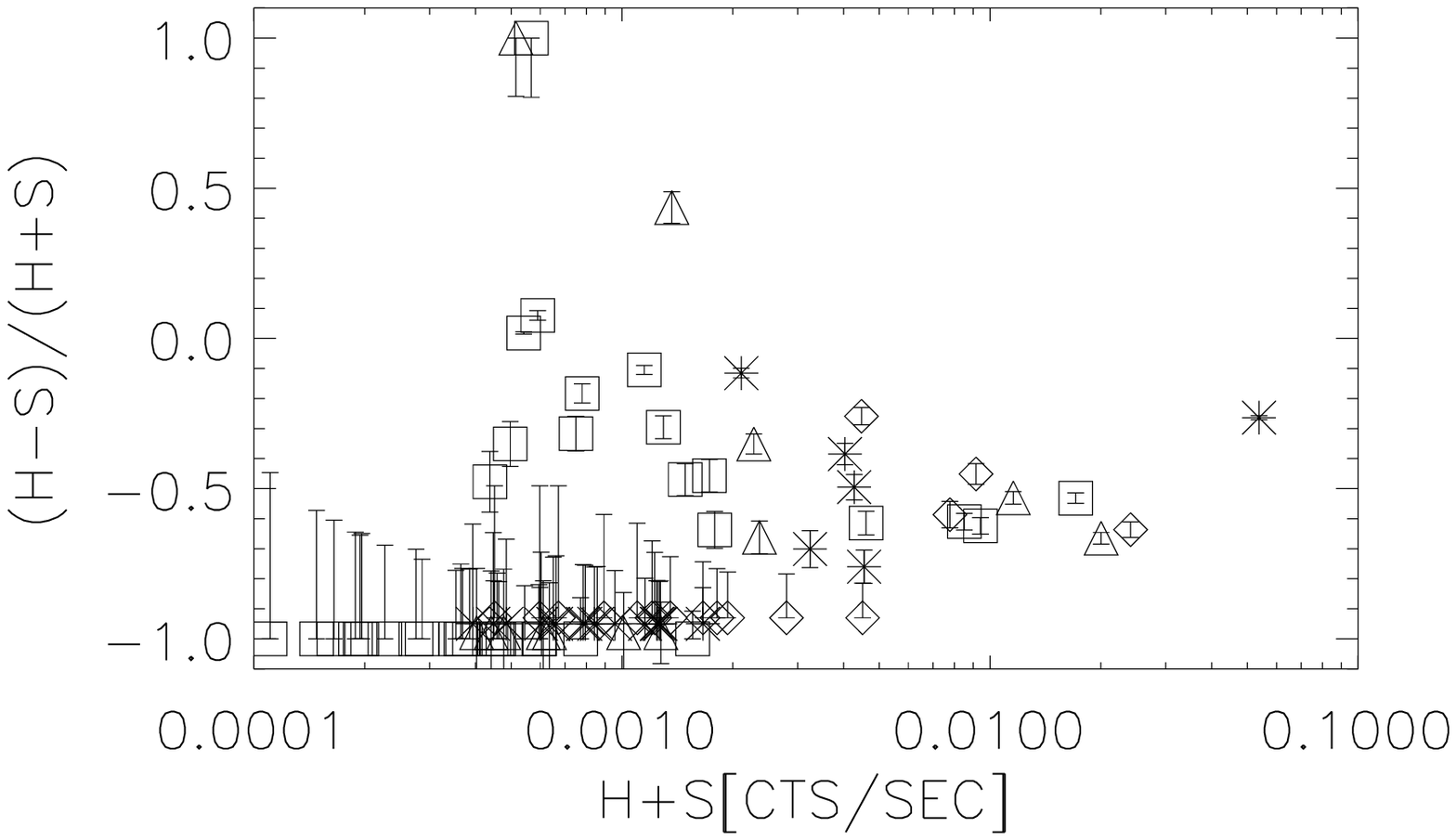}}
\figcaption{ \label{F:FIG2}   
Hardness ratios of unresolved X-ray sources as a function of 0.5-10 keV count rate
in the A1995, MS0451, RXJ0300, and 3C295 fields. 
The symbols are the same as in Figure 1.
}
\end{onecolumnfigure}

\section{Discussion}
\label{s:Discussion}

The derived luminosity functions enable us to estimate the contribution
of the unresolved sources to the overall X-ray emission around clusters
of galaxies. The derived surface brightness of unresolved X-ray 
sources in the $8\arcmin \times 8\arcmin$ fields near RXJ0030 and 3C295 
is comparable to the surface brightness of the X-ray background 
($2.6 \times 10^{-8}\flux str^{-1}$ in the 0.5-2 keV band).
This enhancement is about 100 times larger than the
central enhancement from large scale emission (10 $h^{-1}$ Mpc) 
found by Soltan et al. (1996).
Cappi et al.'s results (cf. their Figure 4) show that these unresolved
sources do not extend beyond the clusters much more than about 4$\arcmin$, 
as opposed to the large scale component of Soltan et al., 
which extends out to about 30$\arcmin$ when scaled to the redshifts of 
RXJ0030 and 3C295 ($z \approx$ 0.5).
Therefore it is likely that this component contributes only to the 
compact component found by Soltan et al.

There are a number of possibilities for the nature of these extra sources:
cosmic variance, 
star-burst galaxies, 
a result of gravitational lensing of background objects, or 
an enhanced number density of AGNs/QSOs.
It is unlikely that the extra sources are due to 
cosmic fluctuations, which is only at the level of about 20$\%$-30$\%$, 
significantly below the measured factor of two (Cappi et al. 2001).
It could be possible, however, that the excess of unresolved sources
is due to projection effects, with differences arising from whether we are
viewing along or perpendicular to a filament of the cosmic web.
Star-burst galaxies are also unlikely to be the sources Cappi et al. found
since their X-ray luminosities are about 10-100 times too faint.
However, recent \CHANDRA deep surveys did find some exceptional 
X-ray bright galaxies at similar redshifts.
Gravitational lensing can increase the number of unresolved sources, 
but only if the \logn slope is steep enough 
($\ge 0.4$, Croom $\&$ Shanks 1999; Mellier 1999).
Two opposite effects are competing in determining the number of observed sources:
lensing magnifies the flux, but it also reduces the field of view behind the
gravitational lens. Lensing would need a significantly higher slope in \logn
to explain the large number of extra sources in the field.
Refregier \& Loeb (1997) predict an average reduction of the surface density
of faint sources at fluxes less than $10^{-15} \flux$. 
Cappi et al. conclude that, as far as spectra and luminosities are concerned, 
the unresolved sources near RXJ0030 and 3C295 could be AGNs/QSOs associated 
with the respective clusters.

In Figure~\ref{F:FIG2} we show the hardness ratios (HRs), $(H-S)/(H+S)$, 
where $S$ and $H$ are X-ray fluxes in the 0.5-2 keV (soft) and in the 2-10 keV 
(hard) bands, of unresolved sources in the fields of A1995 (squares),
MS045 (triangles) as a function of $H+S$ (our results).
As a comparison, we also plot the hardness ratios of RXJ0300 (stars) 
and 3C295 (diamonds) from Cappi et al. (2001).
Points with one sided error bars represent sources not detected either 
in the hard or in the soft band. 
The average HR of unresolved sources detected
in both soft and hard bands near RXJ0300 and 3C295 are $\approx -0.5$, while
the average HR of unresolved sources near A1995 is slightly harder, about $-0.25$.
These sources near A1995 also have lower fluxes than sources near the other two 
clusters, as previously noted (see section 3).

Comparing our Figure~\ref{F:FIG2} to Figure 3 of Rosati et al. (2002), 
which shows the HRs of sources of different types as a function of their 
luminosities, we conclude that unresolved sources in RXJ0300 and 3C295 with 
$H+S \approx 0.004$ cts/sec, corresponding to luminosities of about 
$10^{44} \lum$ at the distance of the clusters, would be compatible to the HRs
of Type I AGNs (as noted by Cappi et al. 2001).
While the faint unresolved sources ($L_* = 5 \times 10^{41}\;\lum$) 
in the field of A1995 are concentrated around $H+S \approx 0.0006$ cts/sec, 
HR $\approx -0.25$, which falls between normal and star-burst galaxies.

Recent results show that the angular correlation function
of X-ray selected AGNs is similar to that of nearby galaxies 
suggesting that AGNs sample the mass density the same way as galaxies sample
(Akylas, Georgantopoulos \& Plionis 2000), in contrast to optically selected 
AGNs, which are found to be more frequent in field galaxies (5$\%$) 
than in galaxies near clusters 
(1$\%$, Dressler Thompson and Shectman 1985; Osterbrock 1960).
Therefore we would expect more X-ray selected AGNs in clusters.
Since the clustering length of X-ray selected AGNs and nearby galaxies is the
same within errors ($\approx 7\;h^{-1}$ Mpc, 
Basilakos 2001; Akylas, Georgantopoulos \& Plionis 2000; Peebles 1993), 
we would expect the ratio of total number of galaxies to the number of 
X-ray selected AGNs, 
$N_{gal}/N_{AGN}$, to be $\approx \langle n_{al}\rangle/\langle n_{AGN}\rangle$
(where $\langle n_{gal}\rangle$ and $\langle n_{AGN}\rangle$ are the average number 
densities of galaxies and AGNs in the cluster).
However, this effect could only account for about 20$\%$ of the excess, much less 
than the a factor of two, which has been found by Cappi et al (2001) and this work.

Our results, that the characteristic luminosities of extra sources are about
one order of magnitude different in A1995 vs. 3C295 and RXJ0030, and that
there seems to be a difference between their HRs argues against cosmic
variance and projection effects.
It suggests instead that different class of objects might dominate in different 
clusters perhaps depending on the formation history and/or the dynamical 
state of the cluster.

Perhaps the unresolved sources in A1995 belong to a class of starburst 
galaxies, a result of enhanced star formation due to interactions between 
infalling groups of galaxies and the intra-cluster gas.
This enhanced star formation would lead to an excess of blue galaxies 
around these areas similar to the Butcher-Oemler effect (Butcher \& Oemler 1978).
A search for a correlation between galaxy color changes around X-ray sources 
compared to other areas in the cluster could be used to check this possibility.

At present, the exact nature of these objects is not known.
Due to limited photon statistic, their spectra could not be determined
individually, both low temperature ($\le 2$ keV) thermal bremsstrahlung and
power low spectra can be fitted to their stacked spectra.
Revealing the physical properties of these objects would help us to improve 
our understanding of structure formation, specifically the origin and
evolution of the intra-cluster gas, and the effect of merging.
Identification of these sources would also help to asses the contamination
these sources cause in the interpretation of cluster emission as 
thermal brems-strahlung from intra-cluster gas.
This contamination would result an overestimation 
of the normalization of the X-ray flux from the cluster 
and would lead to a systematic error in the determination of the
Hubble constant using SZ effect and thermal bremsstrahlung 
(see for example: Molnar, Birkinshaw and Mushotzky 2002).
Follow up observations of the individual sources are necessary to solve 
this mystery.

\acknowledgments 

This work was partly supported by NASA LTSA Grant NAG5-3432 and Chandra 
grant GO0-1049C. The optical analysis was done by Tom Brink. We acknowledge 
useful discussions with Mark Birkinshaw, Renyue Cen, Richard Mushotzky and 
Craig Sarazin.

\end{document}